# A Robust Optimization Approach for Demand Response Participation of Fixed-Frequency Air Conditioners

Jinhua He[a] , Tingzhe Pan[b*], Chao Li[c], Xin Jin[b], Zijie Meng[c], Wei Zhou[c]

[a]*Department of Electrical Engineering, Tsinghua University, Beijing, China;*

[b]*CSG Electric Power Research Institute, Guangzhou, China;*

[c]*Power Dispatching Control Center of Guangdong Power Grid Co.Ltd., Guangzhou, China;*

Tingzhe Pan[b*], corresponding author, E-mail：pantz@csg.cn

# A Robust Optimization Approach for Demand Response Participation of Fixed-Frequency Air Conditioners

With the continuous increase in the penetration of renewable energy in the emerging power systems, the pressure on system peak regulation has been significantly intensified. Against this backdrop, demand-side resources—particularly air-conditioning loads—have garnered considerable attention for their substantial regulation potential and fast response capabilities, making them promising candidates for providing auxiliary peak-shaving services. This study focuses on fixed-frequency air conditioners (FFACs) and proposes an optimization model and solution method for their participation in demand response (DR) programs. First, a probabilistic response model for FFACs is developed based on the Markov assumption. Second, by sampling this probabilistic model, the aggregate power consumption of an FFAC cluster under decentralized control is obtained. Subsequently, a robust optimization model is formulated to maximize the profit of an aggregator managing the FFAC cluster during DR events, taking into account the aggregated response power. The model explicitly considers temperature uncertainty to ensure user comfort in a robust sense. Finally, leveraging the structure of the proposed model, it is reformulated as a mixed-integer linear programming (MILP) problem and solved using a commercial optimization solver. Simulation results validate the effectiveness of the proposed model and solution approach.

## Introduction

With the rapid increase in renewable energy penetration, modern power systems are facing dual challenges: significant fluctuations on both the generation and demand sides, and a lack of system flexibility (Ma 2024). The large-scale integration of intermittent renewable sources, such as wind and photovoltaic (PV) power, has intensified the pressure on peak regulation, while traditional thermal power plants have limited regulation capabilities and poor economic performance. This situation underscores the urgent need to tap into the flexibility potential of demand-side resources. Among these, air conditioning loads—typical temperature-controlled loads—stand out due to their high share (up to 40% of total load in summer), fast response speed, and substantial adjustability potential (Zhu 2016; Zhang 2020). By participating in demand response (DR) programs (Zhang et al. 2013; Chen 2018), air conditioning loads can not only help mitigate the peak-valley gap but also reduce investments in reserve capacity, making them a key focus in optimizing the operation of modern power systems (Waseem 2020). In particular, fixed-speed air conditioners (FSACs) have emerged as highly promising flexible resources for peak regulation services, owing to their widespread adoption and controllable on/off operation (Lou 2018; Ji 2022).

FSACs maintain indoor temperatures through periodic on/off cycling, and their aggregated power consumption can be rapidly adjusted through coordinated control of clusters (Wang et al. 2020; Wang et al. 2018). However, conventional peak regulation strategies often neglect the inherent heterogeneity within air conditioning clusters—such as variations in thermal inertia and user comfort constraints—leading to a lack of load diversity and resulting in power oscillations (Lou 2018; Ji 2022). Furthermore, the decentralized nature of large-scale FSAC clusters, high communication costs, and privacy concerns introduce additional complexities in control implementation (Wang, Zhang and Yao 2019; Hong et al. 2022). Therefore, establishing high-fidelity aggregation models and designing optimization strategies that balance economic performance and system reliability are critical issues in enabling FSACs to participate in DR programs.

Early studies often used the Equivalent Thermal Parameter (ETP) model to describe the dynamic behavior of individual air conditioners (Zhang 2020), but cluster-level modeling must address heterogeneity. Qian et al. 2025 proposed a multi-timescale optimization model for peak regulation, leveraging model predictive control (MPC) to manage uncertainty, yet failed to explore the impact on user comfort in depth. Ma 2024 improved aggregation accuracy through broad-threshold transmission models and secure protocols, but the robustness of its distributed control strategy in real-time task allocation remains to be validated. In contrast, Pei (2018) enhanced large-scale cluster simulation efficiency using Monte Carlo sampling and virtual energy storage modeling, although it did not account for mixed control scenarios involving both fixed-speed and variable-speed air conditioners.

Considering the cyclic nature of FSACs, Ji (2022) developed a batch-based coordinated optimization strategy that effectively mitigates power oscillations. However, its hierarchical distributed architecture relies heavily on a central controller, which imposes significant communication burdens. Wang, Zhang and Yao (2019) proposed a semi-Markov process-based probabilistic control method, enabling virtual peak-shaving unit aggregation through market equilibrium mechanisms, though it lacked a clear user profit-sharing mechanism. Recent studies have shifted toward event-driven and distributed collaborative control. For instance, Hong et al. (2022) introduced an event-triggered consensus control (ETCC) strategy that significantly reduces communication costs, although its convergence proof is limited to linear protocol scenarios.

User comfort is a core constraint in air conditioning-based peak regulation. Wei (2021) developed a Temperature-Humidity Index (THI) model to quantify comfort levels, yet failed to account for individual differences under dynamic environments. Huang (2021) employed neural networks to predict air conditioning loads and designed a tiered control model, but its two-level optimization framework did not incorporate real-time price incentives. Song et al. (2022) combined deep learning with social psychology to quantify user willingness to respond, offering a new approach to potential evaluation, but its parameter identification relied on high-resolution measurement data. In terms of economic feasibility, Guo (2019) proposed a dynamic virtual power plant model, optimizing operation parameters through risk identification, but it lacked effective integration with market clearing mechanisms.

Simulation and experimental validation are essential for verifying the feasibility of theoretical methods. Wang et al. (2018) validated peak regulation strategies on the RT-LAB platform, though its simplified model omitted building thermal inertia. Zhang (2020) verified a compensation pricing model using real-world grid data but did not evaluate the long-term peak regulation performance. More recently, Meng et al. (2024) proposed a direct load control (DLC) optimization strategy for rural building clusters and adopted genetic algorithms for peak shaving (Smith et al. 2013), though its compensation mechanism design remains underdeveloped.

In summary, while substantial progress has been made in modeling, control strategy design, and economic optimization for FSAC-based peak regulation, a trade-off between economic performance and user comfort remains a major challenge. Existing studies often emphasize either economic optimization or comfort modeling (Huang 2021), but the relationship between aggregator profit and user comfort under uncertain outdoor temperatures remains underexplored. This paper aims to maximize the profit of air conditioner aggregators by proposing an optimization model for FSAC clusters participating in DR programs. For the first time, the model explicitly incorporates user comfort robustness to ensure comfort under fluctuating outdoor temperatures.

The remainder of this paper is organized as follows. First, a method for computing the baseline power of FSAC clusters without centralized control is introduced, based on the Markov assumption and implemented using Monte Carlo simulation. This baseline serves as a comparison for controlled scenarios and is used to assess aggregator benefits. Next, the proposed robust optimization model for aggregator profit maximization is presented. Finally, simulation studies are conducted to demonstrate the effectiveness and advantages of the proposed method.

## Materials and methodology

### ***Power Modeling and Monte Carlo Simulation for Aggregated Air Conditioning Clusters Independent of Centralized Aggregator Control Using Markov Chains***

#### *Construction of Markov Chain Model*

Assuming the operational state of the air conditioner follows a discrete-time Markov chain, its state space is $S_t = \{0,1\}$, where a value of 0 denotes the air conditioner being in the "off" state, and a value of 1 represents the "on" state. The specific workflow is illustrated in Figure 1. At any time step t, the state transition probability matrix of the air conditioner is given by Equation (1).

$$p_t = \begin{bmatrix} p_{00} & p_{01} \\ p_{10} & p_{11} \end{bmatrix} \tag{1}$$

$p_{01} = p(S_{t+1} = 0 | S_t = 1)$ denotes the probability of transitioning from the off state to the cooling (on) state, $p_{00} + p_{10} = 1$. $p_{10} = p(S_{t+1} = 1 | S_t = 0)$ represents the probability of transitioning from the cooling (on) state to the off state, $p_{01} + p_{11} = 1$.

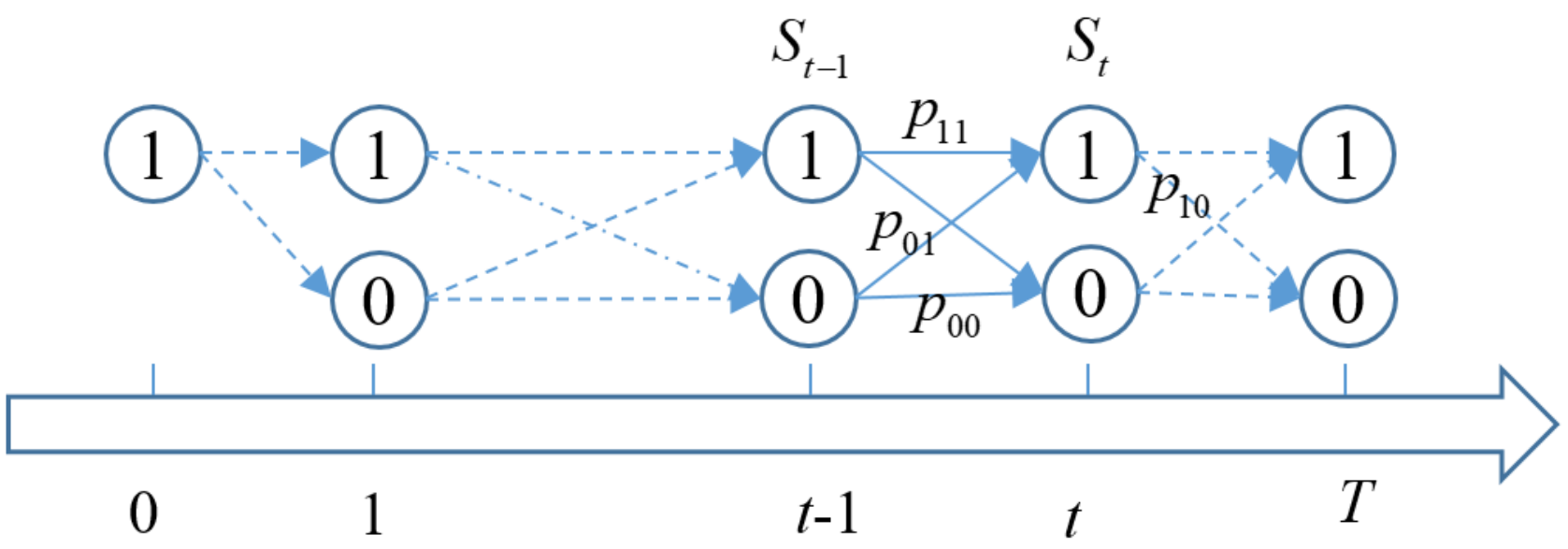

Fig.1. Air conditioner state probability transition process

Assuming the state transition probability of the air conditioner depends solely on the current indoor temperature $\vartheta$, thermostat setpoint temperature $\vartheta_{set}$, and outdoor temperature $\vartheta_{out}$, the state transition probability function can be defined as follows:

$$p_{01} = \sigma\left(a\left(\vartheta_{set} - \vartheta\right) + b\right) \tag{2}$$

$$p_{10} = \sigma\left(c\left(\vartheta_{set} - \vartheta\right) + d\right) \tag{3}$$

Where $\sigma\left(x\right) = \frac{1}{1+e^{-x}}$ is the sigmoid activation function, $\left(a,b,c,d\right)$ represents the model parameters, and in practical engineering applications, $\left(a,b,c,d\right)$ can be estimated via logistic regression.

*Temperature Dynamic Equation*

The thermal dynamics of the air-conditioned room are modeled using a first-order equivalent thermal parameter model, as described in Equation (4).

$$\frac{d\vartheta_t}{dt} = \frac{\vartheta_{out} - \vartheta_t}{RC} + \frac{\eta P_t}{C} \tag{4}$$

Where $R$ denotes the equivalent thermal resistance of the room, defined as the reciprocal of the air heat loss coefficient (unit: °C/kW); $C$ represents the thermal capacitance of the air within the room (unit: kWh/°C); $\eta$ is the Energy Efficiency Ratio (EER) of the fixed-speed air conditioner; $P_t$ denotes the electrical power consumption of the air conditioner, and its relationship with the operational state of the air conditioner is defined in Equation (5).

$$P_t = S_t P_{\text{rated}} \tag{5}$$

In the equation: $P_{\text{rated}}$ denotes the rated power of the air conditioner.

The time variable in Equation (5) is discretized with a time step size $\Delta t$. Assuming the outdoor temperature remains constant over each $\Delta t$ interval, the indoor temperature at time $t+\Delta t$ is given by Equation (6).

$$\vartheta(t+\Delta t)=\vartheta_{out}(t)-R\eta P_t-\left[\vartheta_{out}(t)-R\eta P_t-\vartheta(t)\right]\cdot e^{-\Delta t/RC} \qquad (6)$$

*Monte Carlo Simulation Algorithm*

Based on Equations (2), (3), and (5), the Monte Carlo simulation algorithm can be employed to simulate the power profiles of aggregated air conditioning clusters independent of centralized aggregator control. The simulation workflow for G air conditioners over a total duration of T time steps is summarized in Table 1.

Table 1. Total power simulation calculation process of the air conditioner cluster without aggregator centralized control

| |
|---|
| Step 1: Acquire parameter $(a,b,c,d)$, room thermal capacitance and resistance $(C_g,R_g)$, outdoor temperature forecast $\vartheta_{out}(t)$, set total simulation duration $T$, time step size $\Delta t$, and number of air conditioners $G$; initialize the operational states $S_{g,0}=1$ of all air conditioners. |
| Step 2: Input the outdoor temperature sequence $\vartheta_{out}(t)$, user-set temperature $\vartheta_{g,set}$, and parameter set $(a_g,b_g,c_g,d_g)$.<br>Step 3: For t=1 to T−1:<br>Calculate the state transition probability matrix $p_{g,t}$<br>Calculate the indoor temperature $\vartheta_g(t+\Delta t)$<br>Compute the air conditioner power $P_{g,t}^{ori}$ and total cluster power $P_{g,t}^{ori}=S_{g,t}P_{g,\text{rated}}$<br>End<br>Step 4: Repeat Step 3 to obtain $N$ samples after $N$ simulation. |

***Optimization Model and Solution Method for Fixed-Speed Air Conditioner Clusters Participating in Demand Response***

Air conditioning load aggregators participate in power grid demand response programs based on their own interests. The overall system schematic is shown in Figure 2. The grid dispatch center determines peak load conditions based on day-ahead load forecasting results, then publishes demand response capacity and corresponding time periods in the form of price signals, with higher prices typically occurring during periods of high load demand. The air conditioning aggregator establishes and solves an optimal response model based on peak shaving requirements and status reports (such as ON/OFF state and temperature parameters) from smart terminals of each air conditioning unit, obtaining ON/OFF state curves for each fixed-speed air conditioner. Finally, each smart terminal executes ON/OFF actions according to the state curves.

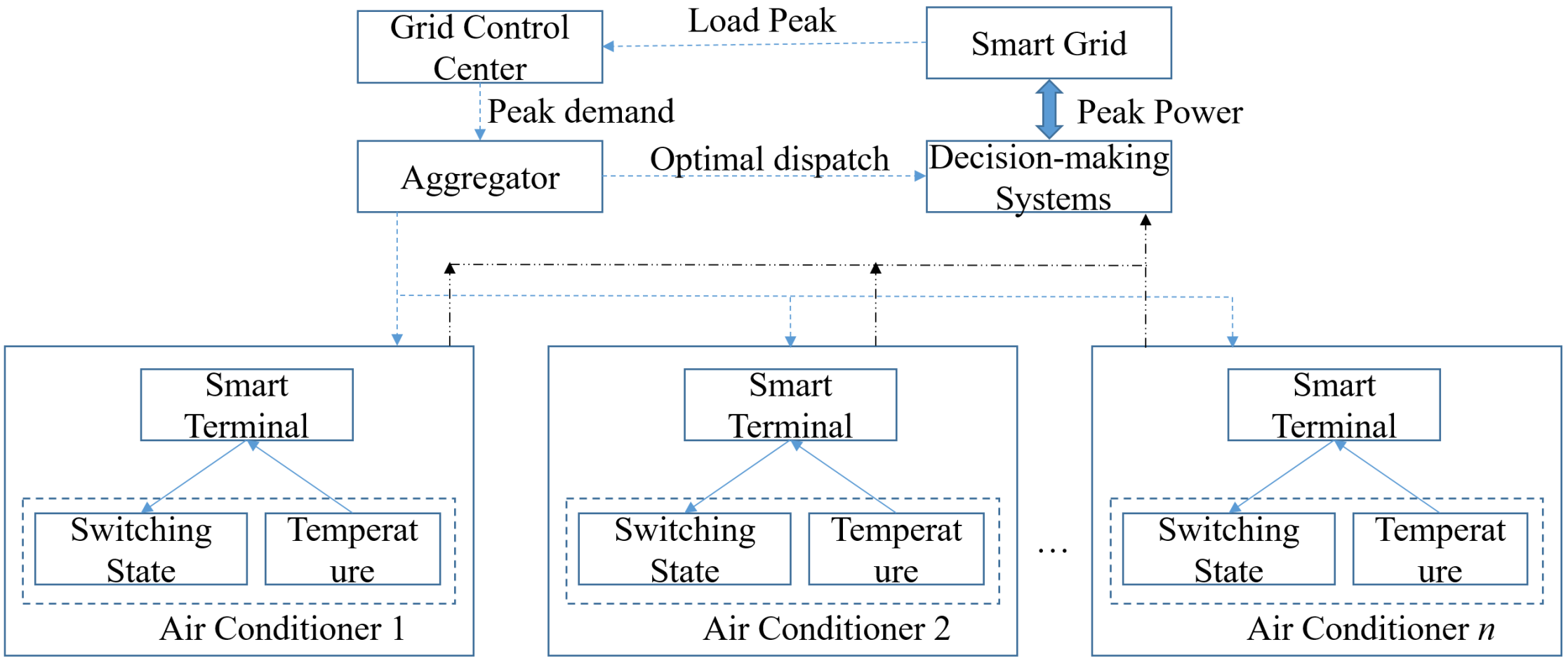

Fig.2. System schematic of the fixed-frequency air conditioner cluster participating in peak shaving services

*Objective Function*

Consider a load cluster consisting of $G$ fixed-speed air conditioners participating in grid peak regulation services over $T$ dispatch periods. The objective is to maximize the benefits of peak regulation by optimizing the on-off control strategy of the air conditioners.

$$\max\left\{\sum_{g=1}^{G}\sum_{t=1}^{T}\left(P_{g,t}^{ori}-P_{g,t}\right)\pi_t\Delta t-\gamma\right\} \tag{7}$$

In the equation: the calculation method of $P_{g,t}^{ori}$ is given in Section 1.3, $P_{g,t}$ represents the air conditioner power, $\pi_t$ denotes the electricity price at time t, and $\gamma$ is the penalty imposed by the air conditioner aggregator for impacts on user comfort.

*Constraints*

1) Since fixed-speed air conditioners can only regulate indoor temperature by switching on/off, their power consumption is a binary discrete value.

$$P_{g,t}=P_{g,\text{rated}}u_{g,t} \tag{8}$$

where: $u_{g,t}$ represents the operating state of the air conditioner (on/off), $P_{g,\text{rated}}$ denotes the rated power of the air conditioner.

2) Thermal Dynamics of Air Conditioners and Rooms

To simplify the model, as described in Section 1.2, this paper adopts a first-order equivalent thermal parameter model for the rooms where the air conditioners are located. The thermal dynamic process of the room can be expressed by Equation (9).

$$\vartheta_{g,t+1}=\vartheta_t^O-R_g\eta_gP_{g,t}-\left[\vartheta_t^O-R_g\eta_gP_{g,t}-\vartheta_{g,t}\right]\cdot e^{-\Delta t/R_gC_g} \tag{9}$$

3) Robust Guarantee of User Comfort

To ensure user comfort, the indoor temperature should not exceed the maximum limit even when outdoor temperatures are high, nor fall below the minimum limit when outdoor temperatures are low.

$$\vartheta_{g,\min}\le\vartheta_{g,t}\le\vartheta_{g,\max}\quad\forall t\ge 2,\forall\vartheta_t^O\in U_t \tag{10}$$

where: $\vartheta_{g,\max}$ and $\vartheta_{g,\min}$ represent the upper and lower bounds of user-tolerable temperature range, $\vartheta_t^O$ denotes the actual outdoor temperature, $U_t$ indicates the uncertainty set of outdoor temperatures.

4) Penalty for User Comfort Impairment

Assuming the user's optimal comfort temperature equals their preset temperature (which remains constant in the short term), any deviation of room temperature from this setpoint reduces comfort level and incurs corresponding penalty.

$$\beta\sum_{g=1}^{G}\sum_{t=1}^{T}\left|\vartheta_{g,t}-\vartheta_{g,set}\right|\Delta t\leq\gamma \tag{11}$$

where: $\beta$ denotes the penalty coefficient, with units of yuan/(°C·h).

5) Initial Temperature Condition

The initial temperature is assumed to be within the comfort range.

$$\vartheta_{g,\min}\leq\vartheta_{g,1}\leq\vartheta_{g,\max} \tag{12}$$

6) Minimum On/Off Duration Constraints for Air Conditioning Equipment

To prevent frequent cycling that may reduce equipment lifespan, minimum duration limits must be enforced for both compressor-on and compressor-off periods.

$$y_{g,t}-v_{g,t}=u_{g,t}-u_{g,t-1} \tag{13}$$

$$y_{g,t}+v_{g,t}\leq 1 \tag{14}$$

$$y_{g,t},v_{g,t},u_{g,t}\in\{0,1\} \tag{15}$$

$$\sum_{q=t-UT_g+1}^{t}y_{g,q}\leq u_{g,t} \tag{16}$$

$$\sum_{q=t-DT_g+1}^{t}v_{g,q}\leq 1-u_{g,t} \tag{17}$$

where: $y_{g,t}$ indicates the air conditioner startup action, $v_{g,t}$ represents the shutdown action, $UT_g$ denotes the minimum unit operating duration, $DT_g$ is the minimum shutdown duration.

7) Uncertainty Set of Outdoor Temperature Variations

Accounting for forecasting errors in weather predictions, the actual outdoor temperature may fluctuate around forecasted values. The uncertainty set for outdoor temperature is shown in Equation (18).

$$U = \left\{ \vartheta^{O} \left| \left\| \vartheta^{O} - \vartheta^{\mathrm{pre},O} \right\| \le \varepsilon \right. \right\} \tag{18}$$

where: $\vartheta^{\mathrm{pre},O}$ represents the predicted value of outdoor temperature. $\vartheta^{O}$ is the random variable, $\|\cdot\|$ is a norm, $\varepsilon$ represents the size of the uncertain set, with larger values indicating stronger uncertainty.

*Constraint Transformation*

In practice, the robustness guarantee for users in the constraint conditions constitutes a robust constraint. Theoretically, this involves an infinite number of constraint conditions that cannot be directly solved using solvers, necessitating an equivalent transformation.

Constraints (9) and (10) are expressed in vector form as follows:

$$\mathbf{Ax} + \mathbf{Bz} + \mathbf{D\xi} \ge \mathbf{h} \quad \forall \xi \in U \tag{19}$$

where: $\mathbf{x}$ represents indoor temperature, $\mathbf{z}$ indicates the on/off status of the air conditioning unit, $\xi$ denotes outdoor temperature, $\mathbf{A}, \mathbf{B}, \mathbf{D}, \mathbf{h}$ represents constant matrices or vectors. Let $\left(\mathbf{x}^*, \mathbf{z}^*\right)$ be the solution satisfying constraint (19). Then $\left(\mathbf{x}^*, \mathbf{z}^*\right)$ is feasible for constraint (19) if and only if constraint (20) holds.

$$\mathbf{e}_j^{\mathrm{T}} \left( \mathbf{Ax}^* + \mathbf{Bz}^* + \mathbf{D\xi} - \mathbf{h} \right) \ge 0 \quad \forall \xi \in U, \forall j \in [m] \tag{20}$$

Constraint (20) holds if and only if the optimal value of the following optimization problem (21) is non-negative for $\forall j \in [m]$.

$$\begin{aligned} &\min_{\xi} \mathbf{e}_j^{\mathrm{T}} \left( \mathbf{A}\mathbf{x}^* + \mathbf{B}\mathbf{z}^* + \mathbf{D}\boldsymbol{\xi} - \mathbf{h} \right) \\ &s.t. \left\| \boldsymbol{\xi} - \hat{\boldsymbol{\xi}} \right\| \le \varepsilon \qquad \left( \theta_j \right) \end{aligned} \tag{21}$$

where $\theta_j$ represents the Lagrange multiplier. Note that the aforementioned problem is a convex optimization problem that satisfies complementary slackness conditions. Consequently, strong duality holds. Based on the Lagrangian duality principle, the dual problem for $\forall j \in [m]$ is given by Equation (22).

$$\begin{aligned} &\max_{\theta_j} \mathbf{e}_j^{\mathrm{T}} \left( \mathbf{A}\mathbf{x}^* + \mathbf{B}\mathbf{z}^* + \mathbf{D}\hat{\boldsymbol{\xi}} - \mathbf{h} \right) - \theta_j \varepsilon \\ &s.t. \left\| \mathbf{D}^{\mathrm{T}} \mathbf{e}_j \right\|_* \le \theta_j \\ &\theta_j \ge 0 \end{aligned} \tag{22}$$

For $\forall j \in [m]$, the optimization problem (21) has a non-negative optimal value if and only if the optimal value of problem (22) is no less than 0. This is equivalent to the existence of $\theta_j$ such that the following constraints (23), (24), and (25) are all satisfied.

$$\mathbf{e}_j^{\mathrm{T}} \left( \mathbf{A}\mathbf{x}^* + \mathbf{B}\mathbf{z}^* + \mathbf{D}\hat{\boldsymbol{\xi}} - \mathbf{h} \right) \ge \theta_j \varepsilon \tag{23}$$

$$\left\| \mathbf{D}^{\mathrm{T}} \mathbf{e}_j \right\|_* \le \theta_j \tag{24}$$

$$\theta_j \ge 0 \tag{25}$$

where $\left\| \cdot \right\|_*$ represents the dual norm. Consequently, by reformulating constraints (9) and (10) as constraints (23) and (25), the original problem maintains equivalence. Thus, the problem to be solved transforms into the following equivalent formulation:

$$\begin{aligned} &\max \left\{ \sum_{g=1}^{G} \sum_{t=1}^{T} \left( P_{g,t}^{ori} - P_{g,t} \right) \pi_t \Delta t - \gamma \right\} \\ &s.t. (8),(11),(12)-(18),(23)-(25) \end{aligned} \tag{26}$$

Problem (26) constitutes a mixed-integer linear programming problem, which can be directly solved using established commercial solvers.

## Results

### ***Parameter Settings***

To verify the effectiveness of the proposed optimal response model for fixed-speed air conditioner clusters participating in peak-shaving services, simulation experiments were conducted based on the frameworks established in Sections 1 and 2. The simulation parameters are configured as shown in Table 2.

Table 2. Parameters of simulation models

| Parameter Name | Symbol | Value/Range | Unit |
|---|---|---|---|
| Number of AC Units | $G$ | 1000 | - |
| Number of Dispatch Periods | $T$ | 48（5-minute intervals, total 4 hours） | - |
| Time Step | $\Delta t$ | 300 | s |
| Rated Power of AC | $P_{rated}$ | * | W |
| Energy Efficiency Ratio | $\eta$ | * | - |
| Thermal Resistance of Room | $R$ | Uniformly distributed in [0.001, 0.00772] | ℃/W |
| Thermal Capacity of Room | $C$ | Uniformly distributed in [336140,3074600] | J/℃ |
| User Set Temperature | $\vartheta_{set}$ | Uniformly distributed in {24，25，26，27，28} | ℃ |
| Outdoor Temperature | $\vartheta_{out}$ | * | ℃ |
| Electricity Price | $\pi$ | * | CNY/kWh |
| Penalty Coefficient | $\beta$ | 3 | CNY/(℃h) |
| State Transition Probability Parameters | $(a,b,c,d)$ | * | * |
| User Tolerance Temperature Range | $\vartheta_{max}, \vartheta_{min}$ | User set temperature $\pm 3$ | ℃ |
| Outdoor Temperature Uncertainty | $\Delta\vartheta$ | $\pm 0.3$ | ℃ |

Among them, the distribution of rated power and energy efficiency ratio for the $G$ air conditioning units is shown in Figure 3, the state transition probability of the air conditioners is illustrated in Figure 4, and the outdoor temperature with time-of-use electricity prices is presented in Figure 5.

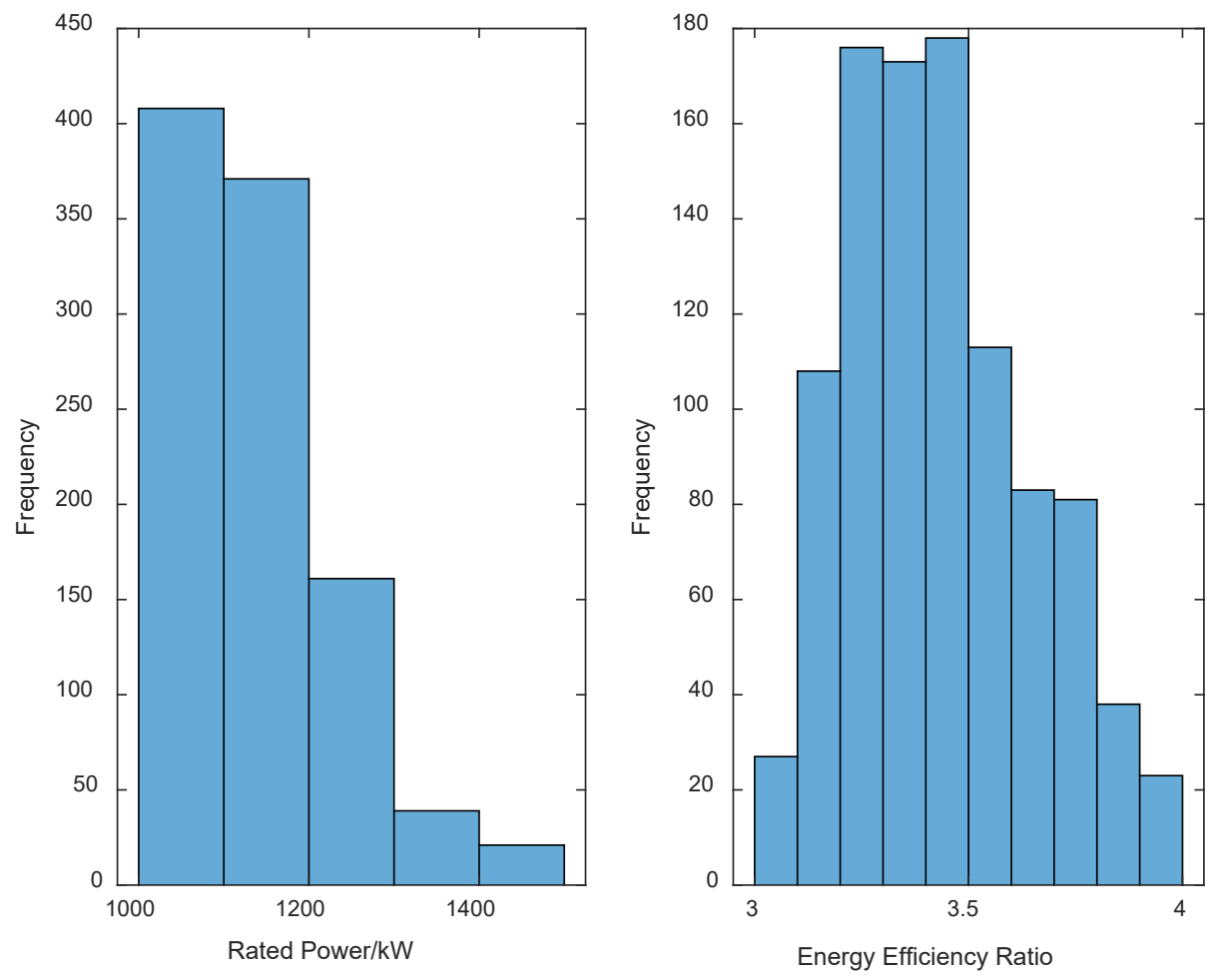

Fig.3 Randomly generated rated power and energy efficiency ratio of fixed-frequency air conditioners: (a) Histogram of rated power distribution, (b) Histogram of energy efficiency ratio distribution

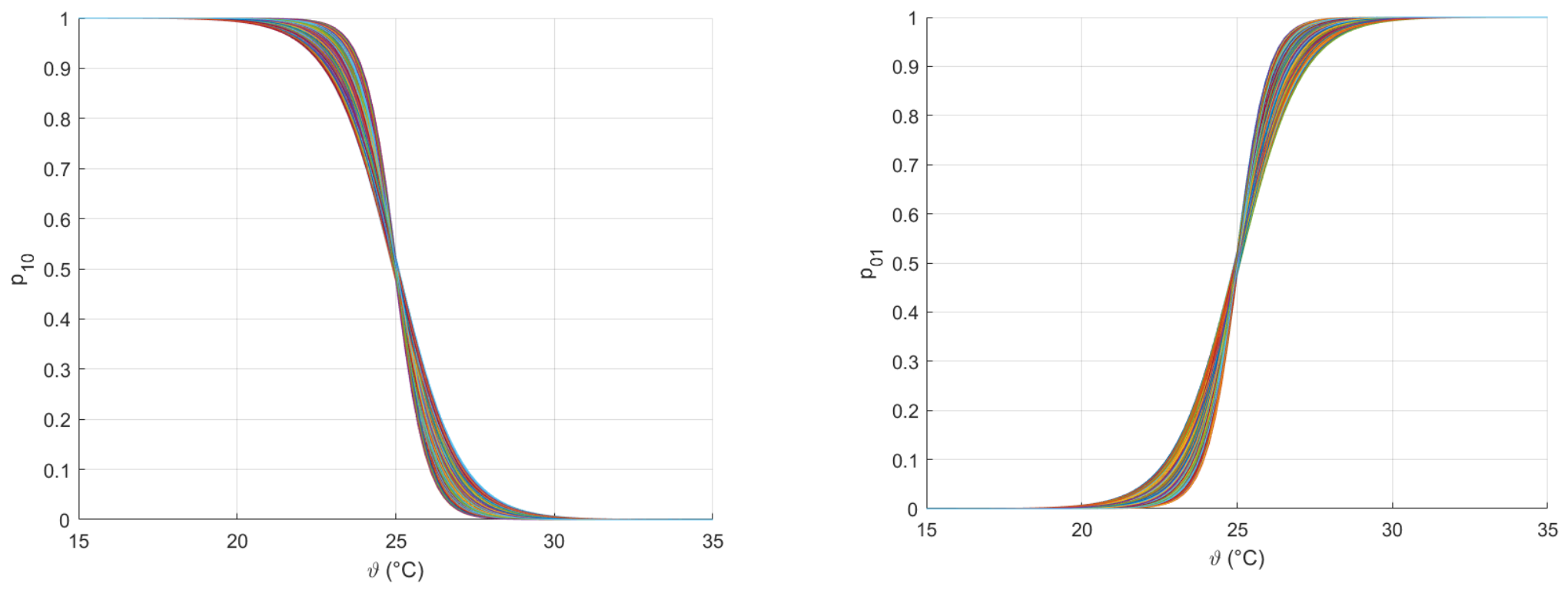

Fig. 4 Curve of air conditioner state transition probability with indoor temperature changes:(a) Relationship between $p_{10}$ and indoor temperature variation, (b) Probability variation with indoor temperature at $T_{set} = 25$

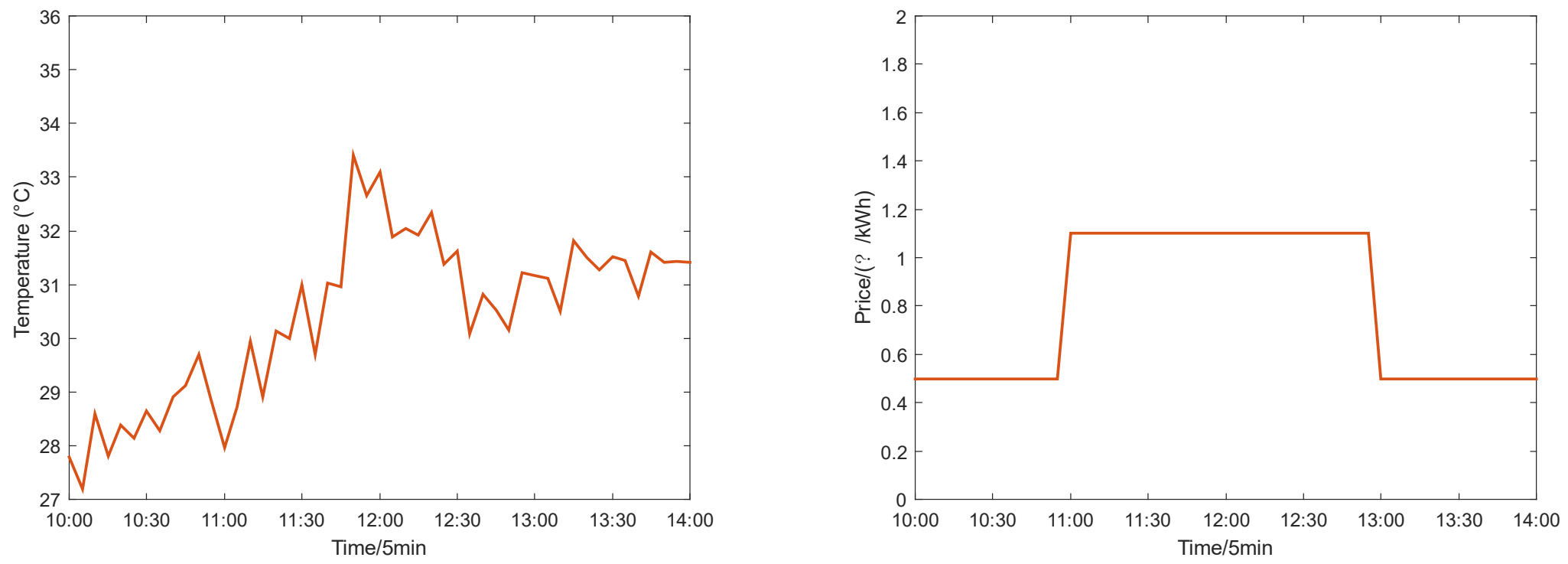

Fig. 5 (a) Outdoor temperature variation curve (b) Time-of-use electricity price curve

*Case Study Results and Analysis*

Based on the above parameter settings, we simulated the power consumption of the air conditioning cluster and indoor temperature changes under both uncontrolled and controlled scenarios, and calculated the revenue of the air conditioning aggregator for both cases.

(1) Uncontrolled Scenario

In the uncontrolled scenario, the cluster's power consumption and indoor temperatures were calculated using a Markov chain model and Monte Carlo simulation. The simulation involved 1,000 air conditioning units with 10,000 sample runs, where the average values across all samples were taken as the cluster's average power and indoor temperature results.

Total Cluster Power: Figure 6 (top) shows the variation of total cluster power over time in the uncontrolled scenario. The initial indoor temperatures followed a uniform distribution $U(25,28)$, while user-set temperatures were uniformly distributed across {24, 25, 26, 27, 28}°C. The cluster's power initially decreased until around noon (12:00), as the outdoor temperature rise was limited, and the cooling capacity of the air conditioners exceeded the heat influx from outdoors, leading to a probabilistic reduction in power consumption. After noon, the total power began increasing due to faster outdoor temperature rise, which increased the probability of users turning on their air conditioners.

Typical User Indoor Temperature: Figure 6 (bottom) displays the indoor temperature variation of a representative user. Before 11:30, the outdoor and initial indoor temperatures were close to the setpoints, resulting in minor fluctuations. Subsequently, as the outdoor temperature rose, the indoor temperature also increased, eventually stabilizing near the user-set value with a deviation of approximately 0.3°C.

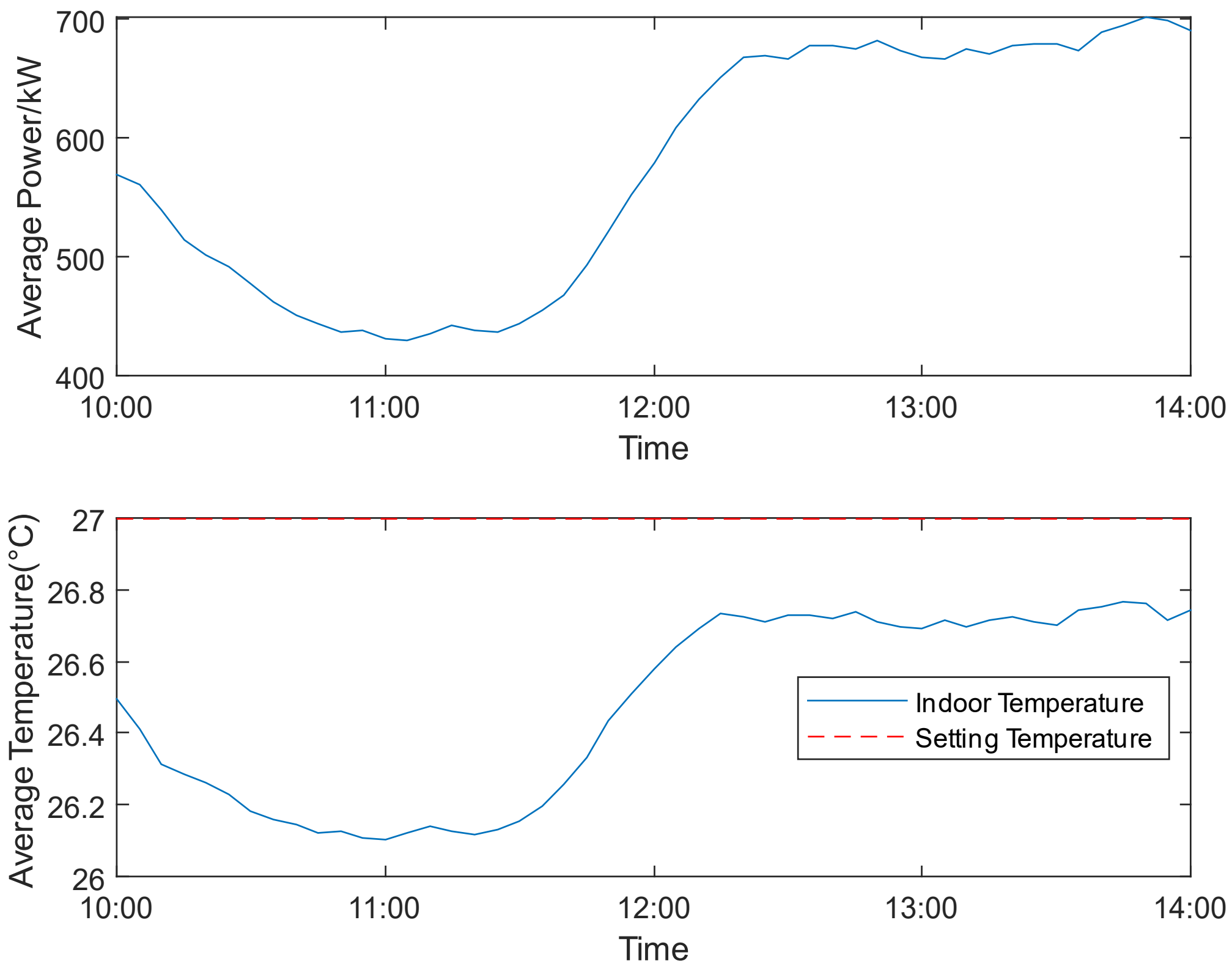

Fig. 6 Total cluster power and typical user indoor temperature in uncontrolled conditions

(2) Controlled Scenario

In the controlled scenario, we implemented the proposed optimal response model for optimized control, aiming to maximize the benefits for the air conditioning aggregator while ensuring robust user comfort guarantees. The mixed-integer linear programming model was solved using the Gurobi solver in MATLAB.

Total Cluster Power: Figure 7 (top) illustrates the variation in total cluster power under controlled conditions. By optimizing the on/off states of fixed-speed air conditioners, the cluster's power consumption during peak electricity price hours (11:00–13:00) was significantly reduced. The maximum power decreased from 0.66 MW (at 12:59) in the uncontrolled scenario to 0 MW (at 12:59) under control, achieving a 100% peak-shaving effect. This demonstrates the model's effectiveness in mitigating grid peak loads. Additionally, the fixed-speed air conditioners fully utilized the thermal storage capacity of user spaces. As shown in the figure, during off-peak hours (before and after the peak pricing period), the cluster's power consumption under control exceeded that of the uncontrolled scenario, indicating proactive cooling operations during these periods.

Typical User Indoor Temperature: Figure 7 (bottom) displays the indoor temperature variation of a representative user under centralized control. Despite fluctuations introduced by optimization, the temperature remained strictly within the comfort range (setpoint ±3°C) at all times.

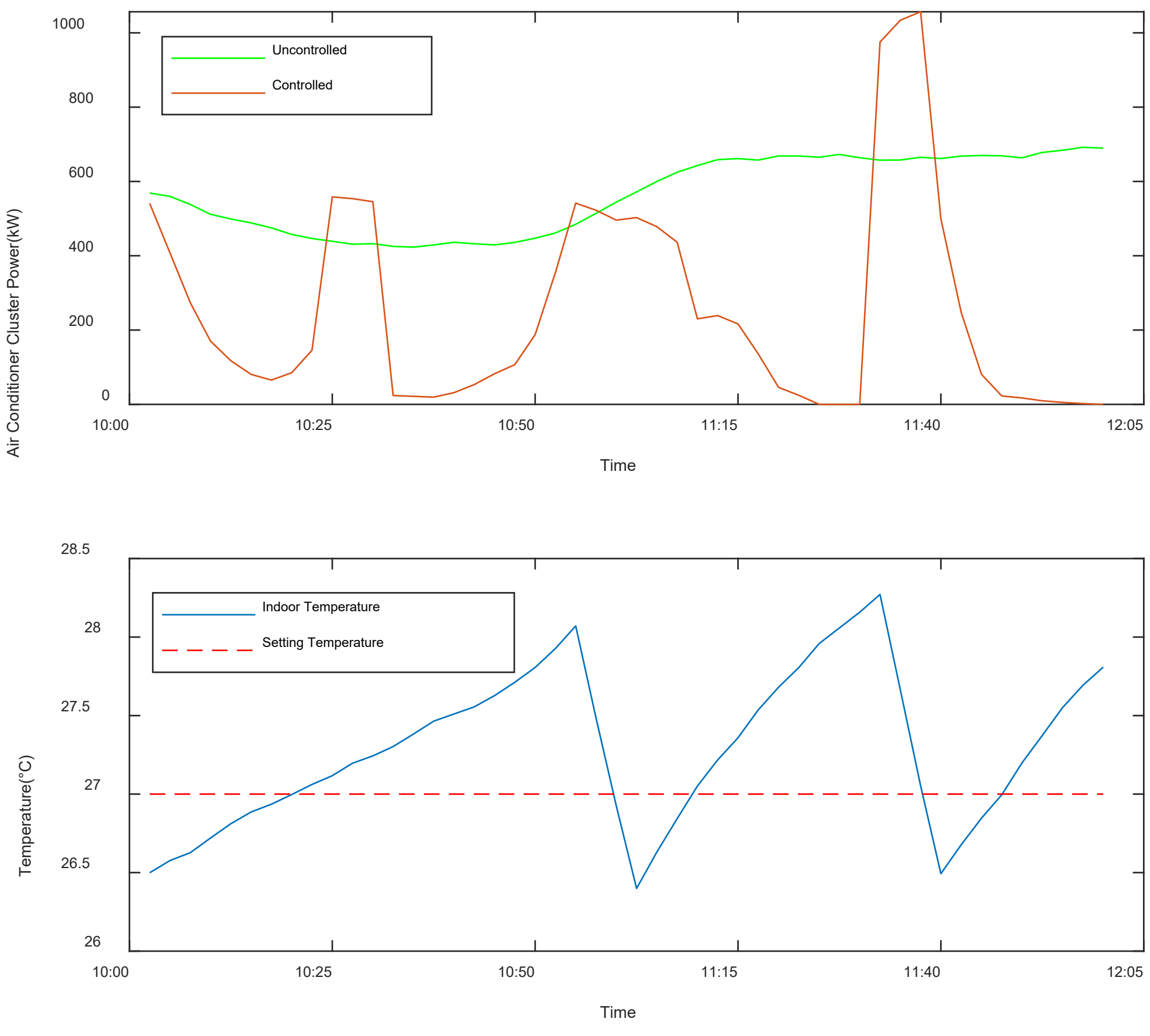

Fig. 7 Total cluster power and typical user indoor temperature in controlled conditions

(3) Revenue of the Air Conditioning Aggregator

Table 3 compares the costs under uncontrolled and controlled scenarios. As shown in the table, the aggregator's revenue is calculated as the difference between the total cost in the uncontrolled scenario and the total cost in the controlled scenario.

Table 3 Cost of the air conditioner cluster in controlled and uncontrolled conditions

| Scenario | Electricity Cost (CNY) | Penalty Cost (CNY) | Total Cost (CNY) |
|---|---|---|---|
| Uncontrolled Case | 1807 | 0 | 1807 |
| Controlled Case | 747 | 812 | 1559 |

As shown in Table 3, under the controlled scenario, electricity costs decreased from 1,807 CNY to 747 CNY, achieving a cost reduction of 1,060 CNY (58.66%). Although temperature deviations introduced a penalty cost of 812 CNY, the total revenue still increased by 248 CNY (13.72%), demonstrating the economic advantages of optimized control.

To further investigate the model's adaptability, we conducted a sensitivity analysis. As presented in Table 4, when the penalty coefficient $\beta$ increases, the model's solution tends to maintain indoor temperatures closer to the setpoints, resulting in reduced revenue for the air conditioning aggregator.

Table 4 Sensitivity Analysis of the Optimal Control Penalty Coefficient for the Air Conditioner Cluster

| $\beta$ /CNY/(°Ch) | 15 | 30 | 45 |
|---|---|---|---|
| Revenue of the Air Conditioning Aggregator / CNY | 728 | 248 | -119 |

## Conclusion

This paper proposes an optimal response model for fixed-speed air conditioning clusters participating in peak-shaving services, with its effectiveness validated through simulation experiments. The key contributions are summarized as follows:

(1) Effective Peak Load Reduction: The optimized control strategy significantly reduces the cluster's power consumption during peak electricity price periods, achieving up to 100% peak-shaving capacity. This not only alleviates grid stress but also enhances power system stability.

(2) Robust Comfort Guarantee: Despite outdoor temperature uncertainties, indoor temperatures are consistently maintained within the comfort range (setpoint ±3°C), rigorously meeting user comfort requirements.

(3) Economic Benefits: The aggregator's participation in peak-shaving services reduces electricity costs and increases total revenue by 13.72%, confirming the model's economic viability.

As thermal resistance/capacity parameters of buildings are difficult to measure accurately in practice, future research will focus on optimization strategies accounting for these parameter uncertainties.